\documentclass[twocolumn,amsmath,amssymb,superscriptaddress,prd]{revtex4}
\usepackage{graphicx}
\usepackage{dcolumn}
\usepackage{bm}

\begin{document}
\title{Modified teleparallel gravity: inflation without inflaton}

\author{Rafael Ferraro}
\email{ferraro@iafe.uba.ar}
\thanks{Member of Carrera del Investigador Cient\'{\i}fico (CONICET,
Argentina)} \affiliation{Instituto de  Astronom\'\i a y F\'\i sica
del Espacio, Casilla de Correo 67, Sucursal 28, 1428 Buenos Aires,
Argentina} \affiliation{Departamento de F\'\i sica, Facultad de
Ciencias Exactas y Naturales, Universidad de Buenos Aires, Ciudad
Universitaria, Pabell\'on I, 1428 Buenos Aires, Argentina}

\author{Franco Fiorini}
\email{franco@iafe.uba.ar} \affiliation{Instituto de  Astronom\'\i
a y F\'\i sica del Espacio, Casilla de Correo 67, Sucursal 28,
1428 Buenos Aires, Argentina}

\begin{abstract}
Born-Infeld strategy to smooth theories having divergent solutions
is applied to teleparallel equivalent of General Relativity.
Differing from other theories of modified gravity, modified
teleparallelism leads to second order equations, since
teleparallel Lagrangian only contains first derivatives of the
vierbein. We show that Born-Infeld-modified teleparallelism solves
the particle horizon problem in a spatially flat FRW universe by
providing an initial exponential expansion without resorting to an
inflaton field.
\end{abstract}


\maketitle

\section{Modified gravity: a Born-Infeld approach}

In 1934 Born and Infeld \cite{Born} proposed the following scheme
for modifying a field theory governed by a Lagrangian density
$\mathcal{L}=\sqrt{-g}\, L$:
\begin{equation}\label{scheme}
\mathcal{L}\longrightarrow\mathcal{L}_{BI}=\sqrt{-g}\, \lambda\,
\left[\sqrt{1+\frac{2\, L}{\lambda}}-1\right].
\end{equation}
The basic idea was to introduce a new scale $\lambda$ with the aim
of smoothing singularities. The scheme (\ref{scheme}) is
essentially the way for going from the classical free particle
Lagrangian to the relativistic one; in such case, the scale is
$\lambda=-m c^2$, which smoothes the particle velocity by
preventing its unlimited growing. Besides, Born and Infeld
subtracted the ``rest energy" to get that $\mathcal{L}_{BI}$
vanishes when $\mathcal{L}$ is zero. We can then expect that
Born-Infeld dynamics will differ from the original dynamics for
those configurations where $L$ is large. In fact, Born and Infeld
looked for a reformulation of Maxwell's electrodynamics in order
to smooth the divergence of the point-like charge electric field,
and they have succeeded in obtaining a finite self-energy for this
configuration. On the other hand, the original Lagrangian is
recovered if $L<<\lambda$; hence the solutions of both theories do
not appreciably differ in these regions. Nowadays Born-Infeld
Lagrangians have reappeared in developments of string theories at
low energies \cite{Frad, Berges, Met, Tsey, Gaunt, Brech}; they
have also been used in quintessence theories for modeling matter
fluids able to drive both inflation and the present accelerated
expansion \cite{sami}. However, although the subject has received
some attention \cite{deser, feingenbaum, comelli}, no
gravitational BI analogue leading to second order equations was
yet proposed in four dimensions.

A wide variety of modified gravity theories have been considered
in the last decades. For instance, Lovelock Lagrangian is a
polynomial in Riemann curvature which leads to second order
equations for the metric tensor \cite{lovelock}. Nevertheless,
Lovelock Lagrangian only differs from Einstein-Hilbert Lagrangian,
$\mathcal{L}_{\textbf{EH}}[g_{\mu\nu}(x)]=-(16\pi
G)^{-1}\sqrt{-g}\, \mathcal{R}$ ($\mathcal{R}$ being the scalar
curvature), for dimension larger than four. On the other hand
``$f(\mathcal{R})$" theories are being currently studied, mostly
connected with the attempts to explain the cosmic acceleration
without resorting to quintessence models \cite{kerner,
capozziello}. For instance, a $f(\mathcal{R})$ theory could be
obtained by using the Born-Infeld scheme:
\begin{equation}\label{elpostahilbert}
\mathcal{L}= -\frac{1}{16\pi G}\sqrt{-g}\, \lambda\, \left[\sqrt{1
+ \frac{2\, \mathcal{R}}{\lambda}}-1\right].
\end{equation}
However we find this strategy unsatisfactory because: 1) fourth
order dynamical equations will result, since $\mathcal{R}$
contains second derivatives of the metric (a feature that is
common to $f(\mathcal{R})$ theories); 2) this strategy is unable
to smooth black holes, since they have $\mathcal{R}=0$ (then the
scale $\lambda$ could not play any role).

Concerning the first objection, it is well known that the second
derivatives of the metric in Einstein-Hilbert Lagrangian do not
lead to fourth order equations because they only give rise to
surface terms in the action. This characteristic only remains
valid in Lovelock Lagrangians but is lost in $f(\mathcal{R})$
theories.

\section{Teleparallel equivalent of General Relativity}

 In order to built a modified gravity leading to second
order equations in four dimensions, we will start not from
Einstein-Hilbert Lagragian but from the teleparallel equivalent of
General Relativity (\textbf{TEGR}). While General Relativity uses
Levi-Civita connection (curvature but no torsion), teleparallelism
uses Weitzenb\"{o}ck connection \cite{Weitz} (torsion but no
curvature). In this sense teleparallelism \cite{albert} is a
sector of Einstein-Cartan theories \cite{cartan, hehl}, which
describe gravity by means of a connection having both torsion and
curvature. In teleparallelism the dynamical object is the vierbein
field $\{\textbf{h}_{i}(x^\mu)\}$, $i=0,1,2,3$. Each vector
$\textbf{h}_{i}$ is described by its components $h_i^\mu$,
$\mu=0,1,2,3$, in a coordinate basis. The matrix $(h_i^\mu)$ is
inversible; i.e. there exist a matrix $(h^i_\mu)$ fulfilling
\begin{equation}
h_i^\mu\ h_\mu^j = \delta^j_i\ ,\ \ \ \ h_i^\mu\ h_\nu^i =
\delta^\mu_\nu .
\end{equation}
Weitzenb\"{o}ck connection,
\begin{equation}\label{Wei}
{\Gamma\!\!\!\!^{^{^{\textbf{w}}}}} {^\lambda_{\mu\nu}} =
-h^i_\mu\,
\partial_\nu h^\lambda_i = h^\lambda_i\, \partial_\nu h^i_\mu ,
\end{equation}
is such that the Weitzenb\"{o}ck covariant derivative of a vector
${\bf V}=V^i {\bf h}_i=V^i  h_i^\mu\, \partial_\mu$ becomes
\begin{eqnarray}
\nonumber && {\nabla\!\!\!\!^{^{^{\textbf{w}}}}}_\nu
V^\mu=\partial_\nu V^\mu+{\Gamma\!\!\!\!^{^{^{\textbf{w}}}}}
{^\mu_{\lambda\nu}} V^\lambda
\\ &&=\partial_\nu(V^ih^\mu_i)
+{\Gamma\!\!\!\!^{^{^{\textbf{w}}}}} {^\mu_{\lambda\nu}}
V^ih^\lambda_i\, =\, h^\mu_i\partial_\nu V^i .
\end{eqnarray}
Hence a vector ${\bf V}$ will be autoparallel if its components
$V^i=h^i_\mu V^\mu$ are constant.

Weitzenb\"{o}ck connection has zero Riemann curvature and non-null
torsion:
\begin{equation}
T^{\lambda}_{\ \ \mu\nu}={\Gamma\!\!\!\!^{^{^{\textbf{w}}}}}
{^{\lambda}_{\nu\mu}}- {\Gamma\!\!\!\!^{^{^{\textbf{w}}}}}
{^{\lambda}_{\mu\nu}}=h_i^\lambda(\partial_\mu
h_\nu^i-\partial_\nu h_\mu^i) ,
\end{equation}
i.e., $h^i_\lambda\, T^{\lambda}_{\ \ \mu\nu}$ are the components
of the 2-form $d{\bf h}^i$, where $\{{\bf h}^i\}$ is the dual
basis (whose elements have components $h^i_\mu$). The
\textbf{TEGR} Lagrangian is \cite{Maluf,Pereira}
\begin{equation}\label{lagrangianotel}
\mathcal{L}_{\textbf{T}}[h^i_\mu(x)]=\frac{1}{16\pi G}\, h\,
S_\rho^{\ \ \mu\nu}\, T^\rho_{\ \ \mu\nu},
\end{equation}
where $h\equiv det(h^{i}_{\mu})$ and $S_\rho^{\ \ \mu\nu}$ is
given by
\begin{equation}\label{tensorS}
S_\rho^{\ \ \mu\nu}=\frac{1}{2}[K^{\mu\nu}_{\ \ \ \rho}
+\delta^\mu_\rho\, T^{\theta\nu}_{\ \ \ \theta}-\delta^\nu_\rho\,
T^{\theta\mu}_{\ \ \ \theta}].
\end{equation}
In this last equation, the \emph{contorsion} tensor is
\begin{equation}\label{contorsion}
K^{\mu\nu}_{\ \ \ \rho} = -\frac{1}{2}\, (T^{\mu\nu}_{\ \ \
\rho}-T^{\nu\mu}_{\ \ \ \rho}-T_\rho^{\ \ \mu\nu}).
\end{equation}
In (\ref{tensorS}) and (\ref{contorsion}), indexes have been
raised and lowered with the metric
\begin{equation}\label{metric}
g_{\mu\nu}(x)=\eta_{ij}\, h^i_\mu(x)\, h^j_\nu(x),\,
g^{\mu\nu}(x)=\eta^{ij}\, h_i^\mu(x)\, h_j^\nu(x)
\end{equation}
($\eta_{ij}=diag(1,-1,-1,-1)$), so it is $h=(-\det
g_{\mu\nu})^{1/2}$. Note that the vierbein is orthonormal in this
metric:
\begin{equation}
g_{\mu\nu}(x)\, h_i^\mu(x)\, h_j^\nu(x) = \eta_{ij} .
\end{equation}
Moreover, Weitzenb\"{o}ck connection proves to be metric
compatible. It is easy to show that the contorsion equals the
difference between Levi-Civita connection associated with the
metric (\ref{metric}) and Weitzenb\"{o}ck connection:
\begin{equation}\label{conexiones}
{\Gamma\!\!\!\!^{^{^{\textbf{w}}}}}
{^{\lambda}_{\mu\nu}}={\Gamma\!\!\!\!^{^{^{\textbf{L}}}}}
{^{\lambda}_{\mu\nu}}+K^{\lambda}_{\ \ \mu\nu} .
\end{equation}
Taking into account the Weintzenb\"{o}ck connection definition
(\ref{Wei}), this means that
\begin{equation}
{\nabla\!\!\!\!^{^{^{\textbf{L}}}}}_\nu h^i_\mu\, =\,
h^i_\lambda\, K^\lambda_{\ \ \mu\nu} .
\end{equation}
 Eq. (\ref{conexiones}) also
means that Weintzenb\"{o}ck four-acceleration of a freely falling
particle is not zero but it is
\begin{equation}
\frac{d^2 x^\lambda}{d\tau^2}+{\Gamma\!\!\!\!^{^{^{\textbf{w}}}}}
{^{\lambda}_{\mu\nu}}\frac{d x^\mu}{d\tau}\frac{d
x^\nu}{d\tau}=K^\lambda_{\ \ \mu\nu}\frac{d x^\mu}{d\tau}\frac{d
x^\nu}{d\tau} .\end{equation} Thus, the contorsion can be regarded
as a gravitational force which moves particles away from
Weitzenb\"{o}ck autoparallel lines.

Teleparallel Euler-Lagrange equations are
\begin{equation}\label{ecuaciones de campo}
\frac{1}{h}\, \partial_{\sigma}(h\, S_{i}^{\ \ \sigma\rho})-4\pi
G\, j_{i}^{\ \ \rho}=4\pi G\, h_i^\sigma\, T^\rho_{\ \ \sigma},
\end{equation}
where $S_i^{\ \ \sigma\rho}=h_i^\lambda\, S_\lambda^{\ \
\sigma\rho}$, $T_\rho^{\ \ \sigma}=h^{-1}\, h^i_\rho\, (\delta
\pounds_{\it{matter}}/\delta h^i_\sigma)$ is the energy-momentum
tensor of the sources and
\begin{equation}\label{corriente}
j_i^\rho=-\frac{1}{h}\,
\frac{\partial\pounds_{\textbf{T}}}{\partial
h^i_\rho}=\frac{h_i^\lambda}{4\pi G} \left(S_\eta^{\ \ \mu\rho}\,
T^\eta_{\ \ \mu\lambda}-\frac{1}{4}\, \delta^\rho_\lambda\,
S_\eta^{\ \ \mu\nu}\, T^\eta_{\ \ \mu\nu}\right).
\end{equation}
Due to the antisymmetry of $S_{i}^{\ \ \sigma\rho}$, a conserved
current appears:
\begin{equation}\label{conservación}
\partial_{\rho}\left(h\, j_i^\rho + h\,
h_i^\sigma\, T^\rho_{\ \ \sigma}\right)=0,
\end{equation}
so $j_{i}^{\ \ \rho}$ is associated with the vierbein
energy-momentum.

\smallskip

Teleparallel Lagrangian (\ref{lagrangianotel}) suffers from a
defect: it is unable to govern the dynamics of the entire
vierbein. In fact, the equivalence between Lagrangian
(\ref{lagrangianotel}) and Einstein-Hilbert Lagrangian tells us
that Lagrangian (\ref{lagrangianotel}) does not govern the
vierbein but just the metric (\ref{metric}). As a reflection of
this character, Lagrangian (\ref{lagrangianotel}) is invariant
under local Lorentz transformations of the vierbein,
$h^i_\mu(x)\rightarrow\Lambda^{i'}_i(x) h^i_\mu(x)$, modulo
boundary terms. In fact, metric (\ref{metric}) does not change
under this kind of vierbein transformation. But we are searching
for a dynamical theory for the vierbein, which is a geometric
object involving sixteen functions $h_i^\mu$ instead of the ten
metric components. Such a theory would govern not only the metric
but also the torsion \cite{kop}. Then, a dynamical theory for the
vierbein requires a Lagrangian differing from
(\ref{lagrangianotel}). In other words, each vierbein on a given
manifold establishes an orthogonal grid of autoparallel lines
which defines an absolute parallelism of vectors (``a vector ${\bf
V}$ is autoparallel if its components $V^i=h^i_\mu V^\mu$ are
constant"). This absolute parallelism is invariant under {\it
global} Lorentz transformations of the vierbein. In fact
$V^{i'}=\Lambda^{i'}_i\, V^{i}$ will be constant only if
$\Lambda^{i'}_i$ is constant as well. Then, a dynamical theory for
the vierbein should be provided with the same global invariance.
Therefore, Lagrangian (\ref{lagrangianotel}) is not admissible.

\smallskip

Other Lagrangians quadratic in the torsion have also been proposed
to build a dynamical theory for the vierbein. In Ref.
\cite{Hayashi1} a general quadratic theory has been tried by
combining three quadratic pieces, each of them associated with
each of the three irreducible parts of the torsion: vectorial,
axial and traceless-symmetric. The coefficients of the vectorial
and traceless-symmetric pieces are strongly constrained by physics
in solar system. The axial part is dynamically linked to the
antisymmetric part of the energy-momentum tensor (associated with
the intrinsic spin of matter). This ingredient renders the theory
invariant only under global Lorentz transformations of the
vierbein. However the use of the axial term has been questioned
(see Ref. \cite{kop}).

\smallskip

In spite of the above mentioned defect, the structure of the
teleparallel Lagrangian (\ref{lagrangianotel}) is very appealing
because it resembles the structure for a gauge field: it is
quadratic in the torsion $h^\lambda_i d{\bf h}^i$. In particular,
it has only first derivatives of the vierbein field. This feature
can be exploited to build a modified teleparallel gravity leading
to second order dynamical equations. Remarkably, modified
teleparallel gravity will be invariant only under global Lorentz
transformations. Concretely, we are going to use a teleparallel
Lagrangian \`{a} la Born-Infeld:
\begin{equation}\label{gravedadmodificada}
\mathcal{L}_{\textbf{{BI}}}=\frac{\lambda}{16\pi G}\, h\,
\left[\sqrt{1+\frac{2 S_\mu^{\ \ \nu\rho}\, T^\mu_{\ \
\nu\rho}}{\lambda}}-1\right].
\end{equation}

Differing from $\mathcal{L}_{\textbf{{T}}}$,
$\mathcal{L}_{\textbf{{BI}}}$ is not invariant under local Lorentz
transformations of the vierbein. In fact, if such transformation
is applied on $\mathcal{L}_{\textbf{{T}}}$ then a harmless
boundary term will appear. But this boundary term emerging from
$S_\rho^{\ \ \mu\nu}\, T^\rho_{\ \ \mu\nu}$ now remains trapped
inside the square root, so rendering the Born-Infeld-like
Euler-Lagrange equations sensitive to local Lorentz
transformations. The Born-Infeld parameter $\lambda$ in Eq.
(\ref{gravedadmodificada}) tells that the metric for solutions of
modified teleparallel gravity will approach the solutions of
Einstein equations in regions where $S_\rho^{\ \ \mu\nu}\,
T^\rho_{\ \ \mu\nu} << \lambda$.

\section{THE COSMOLOGICAL SOLUTION}

Our aim is to test Born-Infeld modified teleparallelism in a
cosmological framework. For this, we will substitute a solution of
the form
\begin{equation}\label{tetrada}
h^i_\mu =\emph{diag}(N(t), a(t), a(t),a(t))
\end{equation}
in the Euler-Lagrange equations emerging from Lagrangian
(\ref{gravedadmodificada}). The proposed solution implies a metric
(\ref{metric})
\begin{equation}\label{metrica friedmann}
g_{\mu\nu}=\emph{diag}(N^2(t),- a(t)^2,- a(t)^2,- a(t)^2),
\end{equation}
i.e., a spatially flat FRW cosmological model. Then we will use as
source a homogeneous and isotropic fluid; so $T^{\rho}_{\ \
\sigma}=\emph{diag}(\rho,-p,-p,-p)$ in the comoving frame. Of
course, the dynamical equations get more involved than the
GR-equivalent ones (\ref{ecuaciones de campo}). The high symmetry
of the proposed solution renders some of the sixteen equations
trivial. Finally only two independent equations are left: a first
order equation
\begin{equation}\label{n(t)=1}
\Big(1-\frac{12\, H^2}{N^2\lambda}\Big)^{-\frac{1}{2}}-1\ =
\frac{16 \pi G}{\lambda}N^2 \rho,
\end{equation}
which results from varying with respect to $h^0_0$
($H(t)=\dot{a}(t)/a(t)$ is the Hubble parameter), and a second
order one,
\begin{equation}\label{ecuacion para a(t)}
\Big(\frac{16 H^2}{N^2\lambda} + \frac{4 H^2}{N^2\lambda}\,
q-1\Big)\Big(1-\frac{12 H^2}{N^2\lambda}\Big)^{-\frac{3}{2}}+1 =
\frac{16 \pi G}{\lambda}\, p\ ,
\end{equation}
which results from varying with respect to $h^i_\sigma$ with $i=$
$\sigma=$1, 2 or 3 ($q = -\ddot{a}\, a/\dot{a}^2$ is the
deceleration parameter). Actually Eqs. (\ref{n(t)=1}) and
(\ref{ecuacion para a(t)}) could also be obtained by replacing the
proposed  solution (\ref{tetrada}) in Lagrangian
(\ref{gravedadmodificada}) and then varying with respect to $N(t)$
and $a(t)$; this is a typical feature of high symmetry solutions.
Note that $S_\mu^{\ \ \nu\rho}\, T^\mu_{\ \ \nu\rho}=-6\,
H(t)^2/N(t)^2$, so $\lambda$ in (\ref{gravedadmodificada}) will
prevent the Hubble parameter from becoming infinite. As it was
expected, Eq. (\ref{n(t)=1}) is not a dynamical equation for
$N(t)$ but a constraint for $a(t)$ (``initial value equation"), as
a consequence of the fact that $N(t)$ is not a genuine degree of
freedom: $N(t)$ can be absorbed by redefining the $t$ coordinate,
so we will choose $N(t)=1$.

By differentiating Eq. (\ref{n(t)=1}) with respect to $t$ and
combining it with Eq. (\ref{ecuacion para a(t)}), the fluid
energy-momentum conservation is obtained:
\begin{equation}
\frac{d}{dt}\, (\rho\, a^3)\ =\ -p\, \frac{d}{dt} a^3.
\end{equation}
If the fluid is described by the state equation $p\, =\, \omega\,
\rho\,$ then one obtains
\begin{equation}\label{conservacion}
a^{3(1+\omega)}\, \rho\ =\ constant \ =\ a_o^{3(1+\omega)}\,
\rho_o,
\end{equation}
where $a_o$ and $\rho_o$ indicate the present-day values.

Combining Eqs. (\ref{n(t)=1}) and (\ref{ecuacion para a(t)}) it
results
\begin{equation}\label{desaceleracion}
1\, +\, q\, =\,
\frac{3}{2}\frac{\left(1+\omega\right)}{\left(1+\frac{16\pi
G}{\lambda}\rho\right)\left(1+\frac{8\pi G}{\lambda}\rho\right)}.
\end{equation}
In General Relativity ($\lambda\rightarrow\infty$) an accelerated
expansion ($q < 0$) is only possible if $\omega < -1/3$ (negative
pressure). Instead, in Born-Infeld modified teleparallelism an
accelerated expansion can be handled without resorting to negative
pressure; a large density $\rho$ is sufficient:
\begin{equation}\label{rhoinflat}
\frac{32\pi G}{\lambda}\rho\ >\ -3+\sqrt{13+12\, \omega}  .
\end{equation}
Actually, for $\rho\rightarrow\infty$ in (\ref{desaceleracion}),
it is $q\rightarrow -1$ and the expansion becomes exponential.

In a context where the cosmological model had spatial curvature,
Eq. (\ref{n(t)=1}) would define the \emph{critical density}
$\rho_c$ making the universe spatially flat. Therefore, it is
useful to measure the contributions to the density coming from
different constituents as fractions $\Omega_i\ =\ \rho_i/\rho_c$.
In this way, by combining Eqs. (\ref{n(t)=1}) and Eq.
(\ref{conservacion}) we obtain
\begin{equation}\label{valores iniciales}
\Bigg(1-\frac{12\dot{a}^{2}}{\lambda
a^{2}}\Bigg)^{-\frac{1}{2}}-1\, =\, \frac{16\, \pi\, G}{\lambda}\,
\sum_i{\rho_o}_i\, \left(\frac{a}{a_o}\right)^{-3(1+\omega_i)},
\end{equation}which can be rewritten in the form
\begin{equation}\label{ecdiferencial}
\dot{\texttt{x}}^2 +
\mathcal{V}(\texttt{x})=0,\hspace{1cm}\texttt{x}=\frac{a}{a_o},
\end{equation}
$\mathcal{V}(\texttt{x})$ being an effective potential given by
\begin{equation}\label{potencial}
\mathcal{V}(\texttt{x})= \frac{\lambda}{12}\ \texttt{x}^2\,
\Big[\big(1 + \beta_o \sum_i{\Omega_o}_i\,
\texttt{x}^{-3(1+\omega_i)}\big)^{-2}-1\Big],
\end{equation} where $\beta_o\equiv (1-12
H_o^2/\lambda)^{-1/2}-1$, is a constant. The potential is always
negative and vanishes with null derivative when $a\rightarrow 0$,
for any value of $\omega$. Moreover, if $\omega > -1/3$ the
potential will asymptotically approach zero when \texttt{x} goes
to infinity. Instead if $\omega < -1/3$ then $\mathcal{V}$ will be
a decreasing function. More relevant is the fact that
$|\mathcal{V}|$ is proportional to $\texttt{x}^2$ when
$\texttt{x}$ goes to zero, so giving an exponential expansion for
the early universe, as was anticipated. If $\omega > -1$ then the
initial behavior is $a(t)\propto \exp[(\lambda/12)^{1/2} t]$.
Therefore, the Hubble parameter is equal to the maximum value
$H_{\emph{max}}=(\lambda/12)^{1/2}$ at the early stage. Eq.
(\ref{ecdiferencial}) also says that
\begin{equation}\label{hubble}
H(z)^2 = H_{\emph{max}}^2 \, [1-(1 + \beta_o \sum_i{\Omega_o}_i\,
(1+z)^{3(1+\omega_i)})^{-2}],
\end{equation}
where $z=a_o/a(t)-1$ is the redshift. Eq. (\ref{ecdiferencial})
for only one constituent ($\Omega=1$) can be easily integrated to
obtain the evolution in an implicit way:
\begin{equation}\label{resultado}
\, \ln\left[2\, (1+ v)+2\, \sqrt{ v\, (2+ v)}\right] -\, \sqrt{
v^{-1}\, (2+ v)}\ =\mathcal{T},
\end{equation}
being $v = \beta_o\, (a/a_o)^{-3 (1+\omega)}$ and $\mathcal{T}=-3
(1+\omega)\ H_{\emph{max}}\ t$.

\section{CONCLUDING COMMENTS}

\begin{figure}[ht]
\centering
\includegraphics[scale=.29]{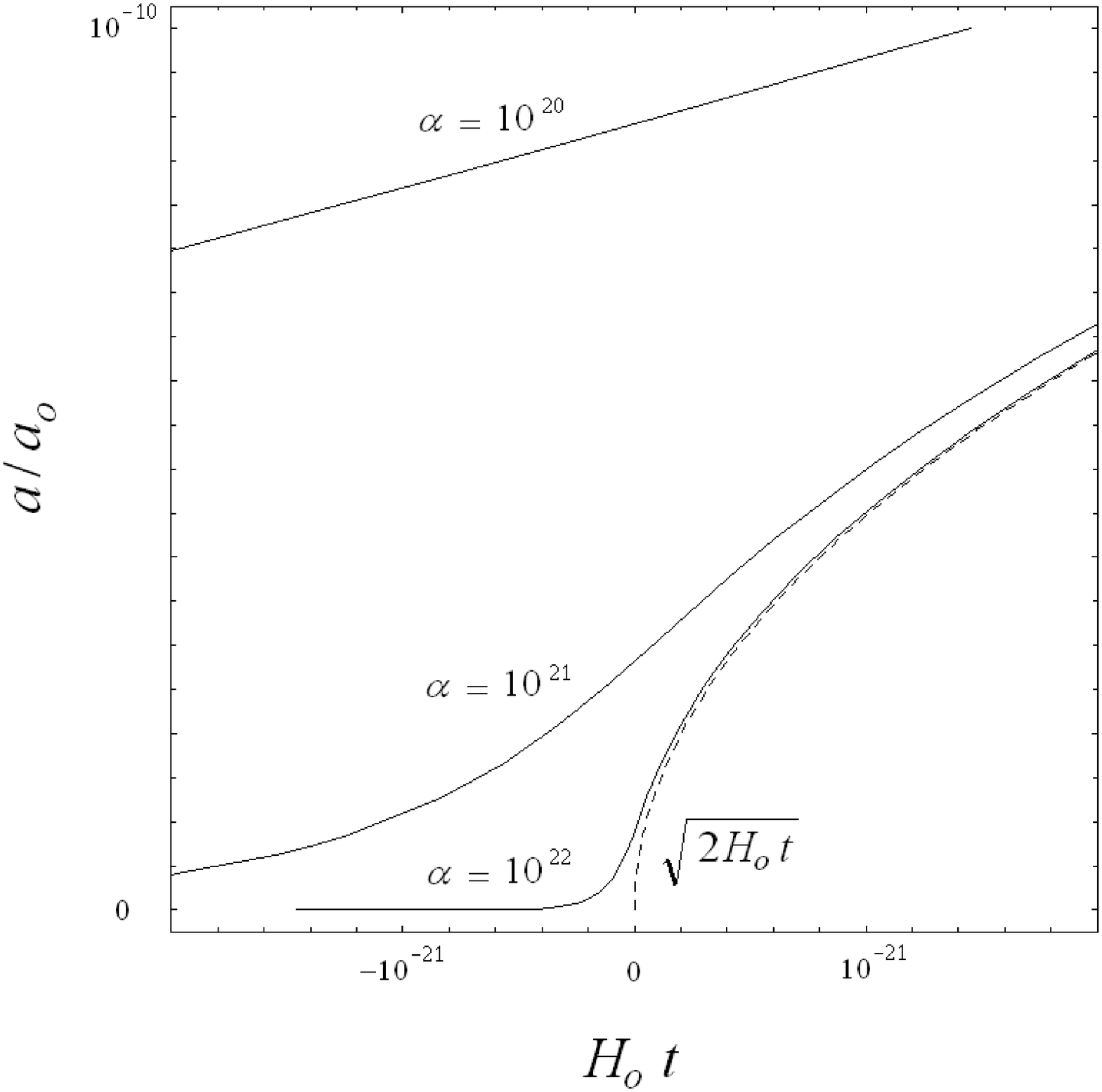}
\caption[]{Scale factor as a function of the cosmological time for
$\omega=1/3$ and different values of $\alpha=H_{max}/H_o$. The
dashed line represents the {\bf GR} behavior. }\label{factor}
\end{figure}

Figure \ref{factor} shows the dimensionless scale factor
$a(t)/a_{0}$ as a function of $H_{0}t$ for several values of
$\alpha=H_{max}/H_o$, as implied by Eq. (\ref{resultado}) with
$\omega=1/3$. The standard ($a/a_{0}=(2H_{0}t)^{1/2}$) behavior is
plotted as a reference (dashed) curve. Remarkably, modified
teleparallelism smoothes the singularity because the scale factor
goes to zero asymptotically.


The main feature of the scale factor behavior is its asymptotic
exponential character for any value of $\omega$. This means that
$H(z)$ becomes a constant when z goes to infinity. This feature
implies that the particle horizon radius
$\sigma=a_{0}\int_{0}^{a_{0}}(a\dot{a})^{-1} da$ diverges. Hence
the whole space-time ends up being causally connected, in
agreement with the isotropy of the cosmic microwave background
radiation. This fact appears as an essential property of modified
teleparallelism which does not require any special assumption
about the sources of the gravitational field.


\begin{figure}[hb]
\centering
\includegraphics[scale=.27]{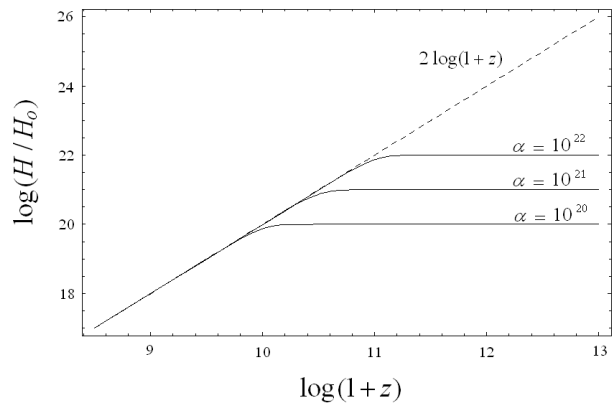}
\caption[]{Hubble parameter as a function of the redshift for
$\omega=1/3$ and different values of $\alpha=H_{max}/H_o$. The
dashed line represents the {\bf GR} behavior. }\label{corrimiento}
\end{figure}

The Standard Big-Bang model successfully explains the relative
abundances of light elements. Therefore, a modified gravity theory
cannot noticeably change the standard evolution of the universe
from the epoch of nucleosynthesis. This means that $H(z)$ at
$z_{nuc}\sim 10^9-10^{10}$ should not appreciably differ from its
standard value. Figure \ref{corrimiento} shows how the Hubble
parameter move away from the {\bf GR} behavior, represented by the
dashed line, to approach the value $H_{max}$ as the redshift
increases. The redshift $z_t$ characterizing the transition
between both behaviors can be defined as the value of $z$ at which
the asymptotic lines intersect. Since the {\bf GR} behavior for
only one constituent is $\log(H/H_o)=(3/2)(1+\omega)\log(1+z)$,
one obtains

\begin{equation}
(1 + z_t)^{3(1+\omega)/2} = \frac{H_{max}}{H_o}
\end{equation}

The condition $z_t >> z_{nuc}$ implies a lower bound for
$H_{max}$. For a radiation dominated universe ($\omega=1/3$) one
obtains that $H_{max}/H_o >> 10^{18}$.

Although inflation without inflaton was already obtained in the
framework of Einstein-Cartan theories (see for instance Ref.
\cite{gasper} and \cite{demian}), those solutions relay on the
existence of spinning matter (the antisymmetric part of the
energy-momentum tensor does not vanish). On the contrary, an
inflationary phase exists in modified teleparallel gravity for a
symmetric energy-momentum tensor. In this case the inflation is
ruled by the parameter $\lambda$ entering the Born-Infeld
Lagragian. $\lambda^{-1/2}$ has dimensions of time, and behaves as
a scale governing the transition from the inflationary phase to
the standard {\bf GR} regime. Besides giving the value of
$H_{max}=(\lambda/12)^{1/2}$, $\lambda$ controls the redshift at
the transition between both regimes.

\acknowledgments

 We are particularly grateful to G. Giribet, M. Leston and M.
Thibeault for so many clarifying discussions. We also would like
to thank C. Simeone and L. Chimento for reading the original
version of the manuscript, and M. Zaldarriaga for helpful
suggestions.

\end{document}